# Solving Configuration Optimization Problem with Multiple Hard Constraints: An Enhanced Multi-Objective Simulated Annealing Approach

Pei Cao

Graduate Research Assistant and Ph.D. Candidate

Department of Mechanical Engineering

University of Connecticut

Storrs, CT 06269, USA

Zhaoyan Fan

Assistant Professor

Department of Mechanical, Industrial and Manufacturing Engineering

Oregon State University

Corvallis, OR 97331, USA

Robert X. Gao

Cady Staley Professor of Engineering

Department of Mechanical and Aerospace Engineering

Case Western Reserve University

Cleveland, OH 44106, USA

Jiong Tang[↑]

Professor

Department of Mechanical Engineering

University of Connecticut

Storrs, CT 06269, USA

Phone: (860) 486-5911, Email: jiong.tang@uconn.edu

---

[↑] Corresponding author

# Solving Configuration Optimization Problem with Multiple Hard Constraints: An Enhanced Multi-Objective Simulated Annealing Approach

Pei Cao[1], Zhaoyan Fan[2], Robert Gao[3], and J. Tang[1][↑]

[1]: Department of Mechanical Engineering
University of Connecticut
Storrs, CT 06269, USA

[2]: Department of Mechanical, Industrial and Manufacturing Engineering
Oregon State University
Corvallis, OR 97331, USA

[3]: Department of Mechanical and Aerospace Engineering
Case Western Reserve University
Cleveland, OH 44106, USA

**Abstract**

This research concerns a type of configuration optimization problems frequently encountered in engineering design and manufacturing, where the envelope volume in space occupied by a number of components needs to be minimized along with other objectives such as minimizing connective lines between the components under various constraints. Since in practical applications the objectives and constraints are usually complex, the formulation of computationally tractable optimization becomes difficult. Moreover, unlike conventional multi-objective optimization problems, such configuration problems usually comes with a number of demanding constraints that are hard to satisfy, which results in the critical challenge of balancing solution feasibility with optimality. In this research, we first present the mathematical formulation for a representative problem of configuration optimization with multiple hard constraints, and then develop two versions of an enhanced multi-objective simulated annealing approach, referred to as MOSA/R, to solve this problem. To facilitate the optimization computationally, in MOSA/R, a versatile re-seed scheme that allows biased search while avoiding pre-mature convergence is designed. Our case study indicates that the new algorithm yields significantly improved performance towards both constrained benchmark tests and constrained configuration optimization problem. The configuration optimization framework developed can benefit both existing design/manufacturing practices and future additive manufacturing.

---

[↑] Corresponding author

**Keywords:** configuration design, multi-objective optimization, Simulated Annealing, hard constraints.

## 1. Introduction

Configuration design and optimization have been studied since the Kepler Conjecture (i.e., no arrangement of equally sized spheres filling space has a greater average density than that of the cubic close packing and hexagonal close packing arrangements). In modern computer-integrated manufacturing, configuration optimizations are frequently encountered in aerospace and automotive systems [1], manufacturing facilities and plants [2-5], and 3-dimensonal laser cutting [6] etc. In general, configuration design involves a wide variety of goals and objectives rather than just optimizing the volume and weight; however, densest packing is a good example in terms of the difficulties one may encounter when dealing with such topics. In 2-dimensional scenarios, one is given a set of geometries such as rectangles, polyminos or spheres. The goal is to pack these items orthogonally into a single rectangular box of unlimited height which needs to be minimized [7], or alternatively, to pack a number of circles inside a circumcircle whose radius needs to be minimized [8]. Three-dimensional problems can be defined in a similar fashion [9], e.g., given a set of three-dimensional objects of arbitrary geometry and an available space (possibly the space of a container), find a placement for the objects within the space that achieves the design objectives, such that none of the objects interferes (i.e. occupy the same space) while optional spatial and performance constraints on the objects are satisfied. The problem of packing has shown to be NP-complete [10], i.e., no optimal algorithm is known running in polynomial time. Therefore, even simple design problems involving spheres, squares or rectangles are known to be difficult problems in the mathematical sense [1]. So far, a global formal solution for such kind of problems has yet to be found, and some believe this will never be accomplished [11]. As such, heuristic approaches are generally required to obtain near-optimal solutions in realistic amount of computational time.

Our work is motivated by the engineering problem of how to spatially arrange a number of components (e.g., cylinders) within a highly restricted space in order to minimize the envelope volume of the assembly as well as the distance between these components that have connective relations (e.g., pipelines) while satisfying certain prescribed constraints. Traditionally, this kind of configuration design problem has been carried out manually based on empirical knowledge due to a large number of constraints brought by manufacturing requirements, which however is time-consuming and does not necessarily yield optimal result. Meanwhile, with the emergence of additive manufacturing in recent years, the components can be manufactured as designed and the connections can be unconventional, which offers a new opportunity to truly facilitate an integrated process of design automation and manufacturing optimization. The form of configuration optimization problems varies for specific applications. For example, as will be shown later the 3-dimensional problem we are trying to address is a cylinder-based problem that has multiple objectives and constrains. Nevertheless, different configuration optimization problems could be abstracted, in a similar manner, as global optimization problems which are generally nonlinear and multi-modal.

Throughout the years, various global optimization strategies have been tailored to address the configuration optimization problems. Dyckhoff [12] used branch and bound approach to solve simple rectangular layout problems with a small number of objects. Szykman and Cagan [13] presented a Simulated Annealing (SA) based approach to generate configurations. Although theoretically general, their work was limited to simple geometric shapes and restricted rotations. Then the work was extended to configuration problems of arbitrary shapes by considering the spatial constrains between components using the same SA approach [14]. Meanwhile, Sachdev et al [15] proposed a framework to integrate techniques such as Simulated Annealing and Genetic Algorithms (GA) to solve configuration design problems. Grignon and Fadel [1] applied multi-objective GA towards configuration optimization problems with more complex objectives. Later on, Aiello et al [16] employed a multi-objective constrained genetic algorithm to solve facility layout problems. A recent review of packing and configuration design methods can be found in [17]. It is worth noting that in previous investigations, the feasibility of the design, i.e., finding a configuration that satisfies constraints, has not been a major challenge. On the other hand, many configuration design problems in modern engineering practices are subjected to multiple hard constraints derived from practical requirements, e.g., the distance between certain components is restricted, and neither overlap nor out-of-bound is allowed. In these scenarios, finding a configuration that satisfies these constraints itself may become computationally demanding. Intuitively, constraints satisfaction should be prioritized in optimization along besides convergence and diversity. So far, there has been no well-established approach to deal with above-mentioned challenges especially for cases with large number of degrees of freedom.

In this research, our goal is to establish a new, systematic framework to tackle the type of configuration optimization problems with multiple hard constraints. We first formulate a compact mathematical model for a representative configuration design problem with many degrees of freedom and hard constraints, and then develop solution algorithms to search for the global optima in realistic amount of iterations. Based on the characteristics of the problem, investigations into heuristic multi-objective optimization algorithms and the treatment of hard constraints are conducted, followed by the development of an enhanced multi-objective simulated annealing algorithm called MOSA/R. The novelty of this algorithm lies in the newly designed re-seed scheme which enables the algorithm to solve the configuration optimization problem as a multi-objective optimization problem much more efficiently than existing algorithms. The rest of the paper is organized as follows. In Section 2, a comprehensive mathematical model of the problem is formulated, where the unique features of such optimization problem are highlighted. We will recognize that it is essentially a constrained non-differentiable multi-objective optimization (MOO) problem. Subsequently, Section 3 gives an overview of representative techniques to solve constrained MOO problems, followed by discussions on their respective pros and cons. A new algorithm, referred to as Multi-Objective Simulated Annealing based on Re-seed (MOSA/R), is then developed in Section 4 to specifically solve the problem. Implementation details and numerical tests results are presented in Sections 5 and 6 respectively, where performance metrics are defined and the performance of two versions of MOSA/R is evaluated and compared to some well-known multi-objective algorithms (MOEA/D, NSGA-II and AMOSA) when applied to benchmark tests and the configuration optimization problem.

## 2. Problem Definition
### 2.1 Case setup

Without loss of generality, we employ one configuration optimization test problem throughout this research. Here we study an optimization problem that involves six cylinders, which are viewed as six functional units that need to be placed within restricted space, with full rotational and translational degrees of freedom. Free-form object representation is precluded to avoid inefficient interference computation. Accordingly, we assume all the cylinders are confined within a cubic space that has a side length of *SL* in a Cartesian system and each side of the cubic space parallels to one of the axes. The surface of the envelope is defined by the largest coordinate value that the cylinder bodies can reach. Note that the surfaces of the cubic space are not the surfaces of the assembly envelope. A representative cylinder is illustrated in Figure 1. One example configuration is depicted in Figure 2. The dimension of each cylinder is defined by its diameter 2*r* and length (Table 1). The connectivity relations of these cylinders are provided in Table 2. The design objectives of this test case are: optimizing the volume of the envelope, meanwhile optimizing the distance of windows that have a connective relationship.

In addition to the aforementioned design objectives, we are also given the following constraints:

*Constraint 1*: Cylinder 1 base must contact cube surface 'a';
*Constraint 2*: Cylinder 6 base must contact cube surface 'b';
*Constraint 3*: Envelope must be located inside the cube, i.e., no cylinder is allowed to surpass the cuboid surfaces;
*Constraint 4*: Connective line 3 must be 5 inches maximum;
*Constraint 5*: Connective line 4 must be 3 inches maximum;
*Constraint 6*: There must be at least 0.5 inch spacing between cylinders.

The six-cylinder model we use in this paper is extracted from an engineering design problem. As each cylinder is allowed to move and rotate, the optimization problem has large design space. Moreover, a critical challenge here is that within such a large design space, the feasible region of design is actually small, because the constraints are demanding and would be easily violated. This test problem can represent a type of configuration optimization problems featuring large design space and constraints that are hard to satisfy. In the numerical experiments, we will vary the boundary constraints (i.e. *SL*) so that different level of constraints is considered.

## 2.2 Mathematical formulation of the optimization problem

As the diameter and length of each cylinder have been specified, five decision variables $x$, $y$, $z$, $\theta$ and $\varphi$ are used to designate the exact position and direction of each cylinder, where ($x$, $y$, $z$) represents the absolute location of the center of cylinder base in the Cartesian coordinate system, and ($\theta$, $\varphi$) represents the direction of the cylinder using spherical coordinate notations (Figure 3). With pre-specified parameters $r$ and $l$, a cylinder can be fully described by ($x$, $y$, $z$, $\theta$, $\varphi$, $r$, $l$). Accordingly, the coordinate of the center of cylinder end can be expressed as

$$\bar{x} = x + l\sin\theta\cos\phi, \quad \bar{y} = y + l\sin\theta\sin\phi, \quad \bar{z} = z + l\cos\theta \qquad (1\text{a-c})$$

Assume $X_e$, $Y_e$, and $Z_e$ are the lengths of envelope sides that parallel with the *X*- , *Y*-, and *Z*-axis, and $d(i, j)$ is the Euclidean distance between the midpoint of axis of Cylinder *i* and Window 1 of Cylinder *j* (i.e., the approximated length of connective line between Window 2 of Cylinder *i* and Window 1 of cylinder *j*). The term 'window' here refers to a specific location on the cylinder that can serve as either the starting point or the destination of a connective line (Figure 1). The *optimization objectives* are

$$\text{Min (Envelope Volume)} = X_e \cdot Y_e \cdot Z_e \qquad (2)$$

$$\text{Min (Connective Line Length)} = \sum_{id=1}^{6} d(i,j)_{id} \qquad (3)$$

Assume the maxima and the minima that the body of Cylinder *k* can reach in the *X*-, *Y*-, and *Z*-axis directions are $x_k^{\max}$, $x_k^{\min}$, $y_k^{\max}$, $y_k^{\min}$, $z_k^{\max}$ and $z_k^{\min}$ respectively. We have

$$\begin{cases} x_k^{\max} = x_k + r_k\sin\theta_k\cos\phi_k \\ x_k^{\min} = \bar{x}_k - r_k\sin\theta_k\cos\phi_k \end{cases} \text{when } x_k \geq \bar{x}_k$$

$$\begin{cases} x_k^{\max} = \bar{x}_k + r_k\sin\theta_k\cos\phi_k \\ x_k^{\min} = x_k - r_k\sin\theta_k\cos\phi_k \end{cases} \text{when } x_k < \bar{x}_k$$

$$\begin{cases} y_k^{\max} = y_k + r_k\sin\theta_k\sin\phi_k \\ y_k^{\min} = \bar{y}_k - r_k\sin\theta_k\sin\phi_k \end{cases} \text{when } y_k \geq \bar{y}_k$$

$$\begin{cases} y_k^{\max} = \bar{y}_k + r_k\sin\theta_k\sin\phi_k \\ y_k^{\min} = y_k - r_k\sin\theta_k\sin\phi_k \end{cases} \text{when } y_k < \bar{y}_k$$

$$\begin{cases} z_k^{max} = z_k + r_k \cos\theta_k \\ z_k^{min} = \overline{z}_k - r_k \sin\theta_k \end{cases} \quad \text{when } z_k \geq \overline{z}_k$$

$$\begin{cases} z_k^{max} = \overline{z}_k + r_k \cos\theta_k \\ z_k^{min} = z_k - r_k \sin\theta_k \end{cases} \quad \text{when } z_k < \overline{z}_k \tag{4a-f}$$

Take Equation 4(a) for example. Its physical meaning is that the maxima/minima for Cylinder $k$ in the $X$-axis direction is specified by the $x$ value of the base/end of the cylinder, the radius, as well as the yaw angle and pitch angle (Figure 3) when $x_k \geq \overline{x}_k$. Thus, we are able to express $X_e$, $Y_e$, and $Z_e$ in the following form,

$$X_e = \max_{i=1,\cdots,6}(x_i^{max}) - \min_{i=1,\cdots,6}(x_i^{min})$$

$$Y_e = \max_{i=1,\cdots,6}(y_i^{max}) - \min_{i=1,\cdots,6}(y_i^{min})$$

$$Z_e = \max_{i=1,\cdots,6}(z_i^{max}) - \min_{i=1,\cdots,6}(z_i^{min}) \tag{5a-c}$$

In Equation (3), *id* stands for the index number of connective lines, e.g., $(i, j)_1 = (1,6)$, $(i, j)_2 = (6,2)$, as indicated in Table 2. $d(i, j)$ is given as

$$d(i,j) = \sqrt{[(\frac{x_i + \overline{x}_i}{2}) - x_j]^2 + [(\frac{y_i + \overline{y}_i}{2}) - y_j]^2 + [(\frac{z_i + \overline{z}_i}{2}) - z_j]^2} \tag{6}$$

Next we consider the mathematical expressions of the constraints. We first assume that the maxima and minima of the cube in the $X$-, $Y$-, and $Z$-axis directions are represented by $Cx_{max}$, $Cx_{min}$, $Cy_{max}$, $Cy_{min}$, $Cz_{max}$ and $Cz_{min}$, respectively, which are determined by where we place the cube in the Cartesian coordinate system and $SL$. As we have defined $SL$ to be the length of each side of the cube, the following must be satisfied,

$$SL = Cx_{max} - Cx_{min} = Cy_{max} - Cy_{min} = Cz_{max} - Cz_{min} \tag{7}$$

Constraints 1 and 2 described in Section 2.1 can now be mathematically expressed as

$$z_1 \equiv Cz_{max},$$
$$x_6 \equiv Cx_{max},$$
$$\theta_1 = 180^o, \quad \phi_1 = 0^o$$
$$\theta_6 = -90^o, \quad \phi_6 = 180^o \tag{8a-d}$$

which essentially indicate that the maximum values in the $Z$- and $X$- axis directions that the cylinders can reach are defined by the $z$ value of Cylinder 1 and $x$ value of Cylinder 6 because their bases will be in contact with cube surface 'a' and 'b'. For Constraint 3, as the entire cylinder cannot pass through the cuboid surface, we can derive

$$\max(x_i^{max}) \leq Cx_{max} = x_6, \quad \min(x_i^{min}) \geq Cx_{min}, \quad i = 1, \cdots, 6$$

$$\max(y_i^{max}) \leq Cy_{max}, \quad \min(y_i^{min}) \geq Cy_{min}, \quad i = 1, \cdots, 6$$

$$\max(z_i^{max}) \leq Cz_{max} = z_1, \quad \min(z_i^{min}) \geq Cz_{min}, \quad i = 1, \cdots, 6 \quad (9\text{a-c})$$

For Constraints 4 and 5 which concern the connective line length, we have

$$d(2,4) \leq 5 - a$$

$$d(4,3) \leq 3 - a \quad (10\text{a, b})$$

Since the connective lines are not always the shortest straight lines between windows, in Equations (11a) and (11b), $a$ is a parameter that translates the connective length to Euclidean distance, which is assumed to be 1 here. We define $D(i, j)$ to be the shortest distance between the center axes of Cylinders $i$ and $j$. As Constraint 6 specifies that there must be at least 0.5 inch spacing between cylinders, we have,

$$D(i, j) \geq r_i + r_j + 0.5, \quad i, j = 1, \cdots, 6; \quad i \neq j \quad (11)$$

In summary, the complete optimization problem formulation can be summarized as

    Minimize envelop volume and connective line length (Equations (2) and (3))

    subject to    1) contact constrains (Equation (8), equalities)

                   2) boundary constraints (Equation (9), inequalities)

                   3) connective line constraints (Equation (10), inequalities)

                   4) overlap constraints (Equation (11), inequalities)

This is inherently a multi-objective optimization problem, as we need to deal with two separate minimization objectives. In order to incorporate the above-mentioned inequality constraints into the analysis, we introduce penalty functions such that these constraints can be treated equivalently as additional objective functions. For Constraints 1 and 2 represented by Equations (9a)-(9c), we define a penalty function

$$\begin{aligned} P_1 = &[\max(x_i^{max}) - Cx_{max}]\Theta[\max(x_i^{max}) - Cx_{max}] + [Cx_{min} - \min(x_i^{min})]\Theta[Cx_{min} - \min(x_i^{min})] \\ &+ [\max(y_i^{max}) - Cy_{max}]\Theta[\max(y_i^{max}) - Cy_{max}] + [Cy_{min} - \min(y_i^{min})]\Theta[Cy_{min} - \min(y_i^{min})] \\ &+ [\max(z_i^{max}) - Cz_{max}]\Theta[\max(z_i^{max}) - Cz_{max}] + [Cz_{min} - \min(z_i^{min})]\Theta[Cz_{min} - \min(z_i^{min})] \end{aligned} \quad (12)$$

where $\Theta(a) = \begin{cases} 1 & \text{if } a > 0 \\ 0 & \text{otherwise} \end{cases}$ is the Heaviside function. Mathematically, minimizing $P_1$ will move the corresponding result toward satisfying the original constraints till $P_1 = 0$ which means the constraints are fully satisfied. Similarly, for the remaining constraints expressed as Equations (10) and (11), we also introduce penalty functions as shown below,

$$P_2 = [d(2,4) - 5 + a]\Theta[d(2,4) - 5 + a] + [d(4,3) - 3 + a]\Theta[d(4,3) - 3 + a] \quad (13)$$

$$P_3 = \sum_{i=1}^{6} \sum_{j=1, j \neq i}^{6} [r_i + r_j + 0.5 - D(i,j)] \Theta [r_i + r_j + 0.5 - D(i,j)] \quad (14)$$

Each kind of inequality constraint has thus been translated into one objective function. It is worth noting that the equality constraints are considered in the move routine of the algorithm to be developed. Our goal now becomes to minimize 5 objective functions expressed as Equations (2), (3), (12), (13), and (14). We have a multi-modal optimization problem associated with 24 design variables and 5 discontinuous objectives within a very large design space.

### 3. MOO Approaches and Constraint Handling Techniques

In this section, we investigate existing techniques that could possibly be used to solve the problem formulated in the preceding section. The problem defined involves simultaneous optimization of several incommensurable and potentially conflicting objectives. Intuitively, multi-objective optimization (MOO) could be facilitated by forming an alternative problem with a single, composite objective function using a weighted sum approach. Single objective optimization techniques are then applied to this composite function to obtain a single optimal solution. However, the weighted sum methods have difficulties in selecting proper weighting factors especially when there is no articulated *a prior* preference among objectives. Indeed, *a posteriori* preference articulation is usually preferred, because it allows a greater degree of separation between the algorithm and the decision-making process which also enables the testing process to be conducted independently of the application [18]. Furthermore, instead of a single optimum produced by weighted sum methods, MOO will yield a set of alternative solutions explicitly exhibiting the tradeoff between different objectives. In light of this, significant amount of research has been carried out on solving MOO problems. Here we start from outlining representative categories of methods of MOO.

### 3.1 Overview of MOO methods

In this sub-section we briefly overview some representative MOO methods, which provides the basis for the subsequent discussion of the handling of constraints and the development of the new algorithm.

The most well-known MOO methods are probably the Pareto-based methods that define optimality in a wider sense that no other solutions in the search space are superior to Pareto optimal solutions when all objectives are considered [19-21]. A general Pareto-based MOO problem where *n* objectives are minimized simultaneously is

$$\text{Minimize} \quad \mathbf{y} = \boldsymbol{f}(\mathbf{x}) = (f_1(\mathbf{x}), ..., f_n(\mathbf{x})) \quad (15)$$

where $\mathbf{x} = (x_1, x_2, ..., x_k) \in \mathbf{X}$ and $\mathbf{y} = (y_1, y_2, ..., y_n) \in \mathbf{Y}$. $\mathbf{x}$ is the decision vector of *k* decision variables, and $\mathbf{y}$ is the objective vector. $\mathbf{X}$ denotes the decision space while $\mathbf{Y}$ is called the objective space. When two sets of decision vectors are compared, the concept of dominance is used. Assuming **a** and **b** are decision vectors, the concept of Pareto optimality can be defined as follows: **a** is said to dominate **b** if:

$$\forall i = \{1,2,...,n\} : f_i(\mathbf{a}) \leq f_i(\mathbf{b}) \tag{16a}$$

and

$$\exists j = \{1,2,...,n\} : f_j(\mathbf{a}) < f_j(\mathbf{b}) \tag{16b}$$

Refer to Figure 4. Any objective function vector which is neither dominated by any other objective function vector of a set of Pareto-optimal solutions nor dominating any of them is called non-dominated with respect to that Pareto-optimal set [19, 21]. The solution that corresponds to the objective function vector is a member of Pareto-optimal set. Usually $\prec$ is used to denote domination relationship between two decision vectors (Table 3).

Another class of methods, the decomposition-based methods, is similar to the weighted sum approach. Unlike traditional weighted sum, they employ a set of weighting vectors to decompose an MOO into a set of single-objective sub-problems. Such methods are compatible with population based techniques such as Genetic Algorithm. A well-known example of such application was conducted by Zhang et al [22]. The third class of methods is the indicator-based methods. They are based on metrics measuring the quality and fitness of the solution. One of the most popular indicators is HyperVolume which was introduced by Zitzler and Thiele [23]. But the calculation of HyperVolume is NP-hard itself, and the computational cost may be prohibitive when the number of objectives is large. Lastly, if *a priori* knowledge is available, which is usually not the case for practical applications, goal-attainment and $\varepsilon$ -constraint could also be considered.

**3.2 Constraint handling**

It is worth emphasizing again that the test problem formulated in Section 2 comes with hard constraints. Here 'hard' means the constraints are not only must-satisfy ones but also difficult to be satisfied (which will be illustrated in Section 6 through actual simulations). Handling constraints within MOO is an important topic that deserves attention particularly when dealing with practical problems that have constraints that need to be incorporated into optimizer in order to avoid convergence towards infeasible solutions [24]. Hence, a biased decision-making technique should be involved favoring the constraints. There are several methods of handling constraints in a multi-objective optimization. One technique is to ignore infeasible solutions. This technique tries to ignore infeasible solutions along the optimization process [25], so that only a newly generated feasible solution will be taken into consideration. But in most practical applications, finding a feasible solution is a major problem itself. Another technique is to use penalty functions, where larger penalty parameters could be assigned to objectives with higher priority. A general formulation takes the form of $\phi(\mathbf{x}) = f(\mathbf{x}) \pm \left[ \sum_{i=1}^{n} h_i \times G_i \right]$ where $\phi(\mathbf{x})$ is the new objective function to be optimized, $G_i$ is the function of the constraint $g_i(\mathbf{x})$, and $h_i$ is a positive penalty parameter. The most common form for $G_i$ is $G_i = \max(0, g_i(\mathbf{x}))$. This method relies heavily on the proper selection of the penalty parameters. If inappropriate parameters are chosen, either a set of infeasible solutions or a poor distribution of solutions is likely. Yet it is still not clear how to select parameters scientifically to guide the search towards the most desirable direction. A third type technique can be regarded as systematic constraint handling procedure [26-28] where the solutions are carefully classified into feasible ones and infeasible ones. However, similar to the challenge in the first technique mentioned above, when most solutions are infeasible along the optimization process, this method would have difficulty in finding solutions.

Because the feasible region is significantly smaller as opposed to the infeasible region for problems with hard constraints, the first technique (ignoring infeasible solution) and the third technique (systematic constraint handling) seem inadequate. In addition, due to the difficulty to determine penalty parameters, the second technique (penalty function) appears to be ad hoc. Alternatively, a potentially promising approach is the combined non-domination check suggested by Ray et al [29]. A general constrained multi-objective optimization problem (in the minimization sense) can be expressed as:

$$\text{Minimize } \mathbf{y} = \boldsymbol{f}(\mathbf{x}) = (f_1(\mathbf{x}),..., f_n(\mathbf{x})) \tag{17}$$

$$\text{subject to } \mathbf{c} \equiv g_i(\mathbf{x}) \geq a_i \quad i = 1, 2, ..., q \tag{18}$$

We can then define a combined objective,

$$\mathbf{y}_{combined} = (f_1(\mathbf{x}),..., f_n(\mathbf{x}), c_1, ... c_q) \tag{19}$$

Then, the non-domination check of the combined objective could be made using Equations (16a) and (16b).

The main idea behind is intuitive, e.g., a constrained single-objective optimization problem will be transformed into an unconstrained MOO problem based on Pareto optimality such that no aggregation of objectives or constraints is involved. This method eliminates the need for fine-tuning penalty parameters so one could approach the feasible region in a more efficient way [30]. It is important to keep in mind that a solution that represents a good trade-off of the constraints, but remains infeasible, may not be of interest in this case whereas it would be acceptable in a constraint-free multi-objective problem. For more details, one may refer to [24, 31, 32].

**4. An Improved Simulated Annealing Algorithm: MOSA/R**

The formulation of an improved algorithm to efficiently solve the configuration optimization problem is presented in this section. Figure 5 shows the compatibility relations between constraint handling techniques and multi-objective handling techniques. As we hope to use combined non-domination check to handle constraints, the only approach that is cohesive with the combined non-domination check is the Pareto-based approach. Thus, we decide to employ the combination of Pareto-based approach and non-dominant check to solve the hard constrained problem. Essentially, a combined objective as demonstrated in Equation (19) is used as the design objective followed by the Pareto domination check as illustrated in Equations (16a) and (16b). Therefore, neither weighted aggregation nor feasibility check is involved. Such combination can be potentially incorporated by two kinds of stochastic MOO algorithms, Genetic Algorithms (GA) and Simulated Annealing (SA), which happen to be considered as the effective approaches in solving configuration design problems [17]. In this application, SA is more promising than GA for several reasons. First, the decision space of our test problem is larger compared to test cases like ZDT, DTLZ or WFG in literature [33]. The synergy effect of large decision space and multi-modality may render two sets of approximated optimal solutions sitting very far away from each other in decision space. As suggested by Beyer [34], the crossover operator has the effect of genetic repair which survives the common features of both parents. The crossover operation may not work well towards the test problem because the parents would probably have nothing in common. The second reason is that most solutions we find along the course of optimization would be infeasible. GA, as a population-based algorithm, generates more infeasible solutions that are needed to be discarded than that of SA when executed under the same number of evaluations, which means GA is more computationally intensive (or converges slower) than SA under such circumstance. As reported by Mann and Smith [35], the execution time of GA is from 10 to 24 times longer than that of SA. On the other hand, SA only generates one solution at a time, which is easier to manipulate towards feasibility. In fact, SA can, in a certain way, be viewed as a special case of GA [36]. As will be shown later, our simulation results match the above statement well (see Section 6.1 and Section 6.2). However, for the configuration optimization test problem analyzed in this research, the GAs (MOEA/D and NSGA-II) that we apply cannot find feasible solution in a reasonable time frame. Thus, the corresponding results are not included in Section 6.3.

Simulated Annealing [37] is a heuristic technique drawing an analogy from physics annealing process. It was originally designed for solving single objective optimization problem and was proven to be robust and convergent if annealed sufficiently slow [38, 39]. It is not obvious at first how to extend the implementation to a multi-objective context other than through a weighted sum approach until Engrand, who is among the very first to embed the concept of Pareto optimality with SA, proposed to maintain an external population archiving all non-dominated solutions found so far [40]. Then several Multi-Objective Simulated Annealing (MOSA) methods that incorporate Pareto set [41-46] have been developed. The acceptance criteria of these methods are all derived from the differential between the new and current solutions. However, in the presence of Pareto set, solely comparing the new solution to the current solution appears to be vague. That is why there have been a few techniques proposed that use Pareto domination based acceptance criterion in MOSA [47-52], the salient feature of which is that the domination status of the point is considered not only with respect to the current solution but also the archive of non-dominated solutions found so far. It has been widely demonstrated that simulated annealing algorithms are capable of finding multiple Pareto-optimal solutions in a single run.

**4.1 Algorithm formulation**

Based on the discussion above, we now develop a re-seed based MOSA algorithm, hereafter referred to as MOSA/R, following the lead of AMOSA [50] that uses dominance measure to compute the probability of acceptance of a new point. This new MOSA/R scheme will enable constraints handling without the need to specify parameters through the use of a re-seed combined with non-dominated rank of the constraints. In this research, two different versions of MOSA/R are tested. The feasibility and efficiency of the proposed methods with application to the test problem defined earlier are illustrated and compared with AMOSA. The pseudo-code of MOSA/R is provided below.

**Algorithm MOSA/R**

1. Set *Tmax*, *Tmin*, # of iterations *iter*, cooling rate *α*
2. Initialize the *Archive* (Pareto front)
3. *Current solution* = randomly chosen from *Archive*
4. While (*T* > *Tmin*)
5.    For 1 : *iter*
6.       Generate a *new solution* vector in the neighborhood of *current solution* vector
7.       If *new solution* dominates *k* (*k* >= 1) solutions in the *Archive*  /*Case 1*/
8.          Remove all *k* dominated points from the *Archive*
9.          Add *new solution* to the *Archive*
10.         Set *new solution* as *current solution*
11.      Else if *new solution* dominated by *k* (*k* >= 1) solutions in the *Archive* /*Case 2*/
12.         If *new solution* dominated by *current solution* /*Case 2a*/
13.           If *current solution* ∈ *Archive* /*Case 2a-1*/
14. $$prob = \frac{1}{1+\exp(\Delta dom_{avg}/T)}$$

  where $\Delta dom_{avg} = \dfrac{\sum_{i=1}^{k} \Delta dom_{i,new}}{k}$

15.     Set *new solution* as *current solution* with *prob*
16.    Else if *current solution* $\notin$ *Archive* /*Case 2a-2*/
17.     Select a solution from *Archive* following <u>designed rules</u>
18.     $prob = \dfrac{1}{1 + \exp(-\Delta dom_{select,new})}$

    where $\Delta dom_{select,new}$ = domination amount between *selected solution* and *new solution*

19.     $prob' = \dfrac{1}{1 + \exp(\Delta dom_{avg} / T)}$

    where $\Delta dom_{avg} = \dfrac{(\sum_{i=1}^{k} \Delta dom_{i,new}) + \Delta dom_{current,new}}{k}$

20.     Set *selected solution* as *current solution* with *prob*
21.     Set *new solution* as *current solution* with (1-*prob*)\**prob'*
22.    End if
23.   Else if *new solution* dominates *current solution* /*Case 2b*/
24.    Set *new solution* as *current solution*
25.   Else if *new* and *current solution* are non-dominant to each other /*Case 2c*/
26.    $prob = \dfrac{1}{1 + \exp(\Delta dom_{avg} / T)}$

   where $\Delta dom_{avg} = \dfrac{(\sum_{i=1}^{k} \Delta dom_{i,new})}{k}$

27.    Set *new solution* as *current solution* with probability=*prob*
28.   End if
29.  Else if *new solution* and *Archive* are non-dominant to each other /*Case 3*/
30.   Add *new solution* to the *Archive*
31.   Set *new solution* as *current solution*
32.  End if
33.  End for
34. $T = \alpha * T$
35. End while

Similar to AMOSA, MOSA/R uses the concept of the amount of domination in computing the acceptance probability of a new solution. Given two solutions **a** and **b**, the amount of domination is defined as

$$\Delta dom_{\mathbf{a},\mathbf{b}} = \prod_{i=1, f_i(\mathbf{a}) \neq f_i(\mathbf{b})}^{M} (|f_i(\mathbf{a}) - f_i(\mathbf{b})| / R_i) \tag{20}$$

where $M$ is the number of objectives and $R_i$ is the range of the $i$th objective [50].

**4.2 The re-seed scheme: when it needs to be triggered**

The re-seed scheme (Case 2a-2) (Line 16 to Line 22) essentially differs MOSA/R from AMOSA. A re-seed scheme consists of two parts, "when" to re-seed and "what" to re-seed. "When" implies the scenarios that triggers the re-seed, which is problem independent. And "what" means if the re-seed happens, which solution should be selected. Such selection process could be designed to accommodate different optimization needs. Let's first define when to re-seed in a general sense. (What to re-seed will be discussed in Section 5.3). Figure 6 illustrates when the re-seed may happen for both AMOSA and MOSA/R in the objective space and the decision space. Figures 6(a) and 6(b) depict the simplified scenarios where two objectives are considered while Figures 6(c) and 6(d) are the simplified scenarios with only one objective. In AMOSA, the probability re-seed happens when the *new solution* dominates *current solution*, and at the meantime, is dominated by at least one solution in the *Archive*. As illustrated in Figures 6(a) and 6(c), if re-seed takes place, a potential non-dominated solution (the dotted line circle) could be overlooked, which, as a consequence, will lead to premature convergence, and loss of diversity among the Pareto front. However, in MOSA/R, the probability re-seed happens when a *new solution* is dominated by both *current solution*, which does not belong to the *Archive*, and, at least, one solution in the *Archive*, as depicted in Figures 6(b) and 6(d). The re-seed timing of MOSA/R is improved compared to that of AMOSA in the sense that it gives the algorithm more flexibility to explore a possible optimal region. Figures 6(e) and 6(f) show that given the *current solution* and the *Archive*, how the optimizer performs correspondingly when the *new solution* falls into different quadrants for AMOSA and MOSA/R respectively. It is worth noting that the presence of re-seed inherently may cause a possible loss of diversity or even, at worst, missing the global optimal. Nevertheless, for most practical applications where the objective space is considerably large, a re-seed scheme is preferable due to the fact that it will enable the algorithm to converge to the feasible region in a timely manner. A reasonable convergent time is significant because otherwise exhaustive random search, which could be easily proved to be convergent to optimal if given enough time, would be a panacea. Hence, even though Simulated Annealing has been proved to be convergent, the estimated time of absolute convergence $\|v(kr) - e^*\| = O(1/k^{\min(a,b)})$ which is related to the objectives and annealing schedule [39] is usually unrealistic. Under such circumstances, a properly designed re-seed could make a difference.

The merit of simulated annealing is to allow inferior moves by probabilistic relaxation, so potentially an inferior solution will lead the search to a superior solution in the subsequent steps. The re-seed scheme of AMOSA weakens such merit because once the new solution shows the potential for being the starting point of the very path towards the optimal region, re-seed may be triggered as in the scenarios in Figures 6(a) and 6(c). Table 4 gives a direct comparison of the decision procedure in MOSA/R and AMOSA. The discussion of how to re-seed (i.e., Line 17 designed rules in the pseudo-code) is presented in the next section of implementation.

## 5. Implementation Details
### 5.1 Initialization

The algorithm begins with 100 initial solutions in the *Archive*. All these solutions, which correspond to possible configurations, are chosen randomly from the objective space. These solutions can also be refined using a simple hill-climbing technique as suggested in [50]. In this research, we employ a randomly chosen initial *Archive* to better exploit the performance of each algorithm.

### 5.2 Move routine

According to Line 6 of the pseudo-code, the new solution is obtained through perturbations. For the test problem, perturbation means to move the cylinders. In this research, we define two move routines, i.e., translation and rotation. To be more specific, translation perturbs *x*, *y* and *z*, and rotation perturbs $\theta$ and $\varphi$. The new perturbed decision variable is sampled from Laplace distribution whose expectation is the current solution and diversity parameters are chosen depending on the scale of each variable. The probability density function is given as

$$f(x\mid\mu,l) = \frac{1}{2l}(-\frac{|x-\mu|}{l}) \tag{21}$$

where $\mu$ denotes the expectation and $l$ is the diversity parameter. For our case, $l$ is chosen as 0.5 and 30 for translation and rotation, respectively. We only carry out one kind of perturbation to one cylinder per iteration for the purpose of detailed scanning of the objective space. Nevertheless, it is also viable to perturb multiple cylinders at once.

### 5.3 The re-seed scheme: what solution to pick once re-seed is triggered

We have yet to discuss what solution to select from the *Archive* (Line 17 designed rules) to compete for acceptance. One of the advantages of the selection process is that it could be adjusted flexibly with respect to different applications. In this research, we will formulate two versions. First is to select the solution from *Archive* that corresponds to the minimum difference of domination amount. Then the selected solution is set as a current solution with probability $1/(1+\exp(-\Delta dom_{select,new}))$. MOSA/R using the first approach is given the notation MOSA/R-1.0.

The second approach needs to make use of the technique called fast non-dominated sort [53]. We sort the non-dominated solutions in *Archive* into fronts based on their objective functions that correspond to constraints (Equations (12), (13) and (14)). This is the step that embodies biased constraints handling. Then, the solution in the first front with a minimum amount of domination towards the new solution is chosen and will be set as the current solution with probability $1/(1+\exp(-\Delta dom_{select,new}))$. MOSA/R using the second approach is given the notation MOSA/R-2.0. In both approaches, the solution corresponding to the minimum difference of domination amount is chosen instead of the solution corresponding to the maximum difference of domination to avoid premature convergence. Figure 7 illustrates how to choose one solution among *Archive* as a re-seed candidate when there are two penalties. In our case study, most solutions along the process are infeasible solutions, so the issue of maintaining diversity among population seems somewhat trivial. When a well distributed and well spread non-dominated set is something needs to be achieved, techniques such as fitness sharing [54], isolation by distance [55], over specification [19, 56], crowding [53], $\varepsilon$-dominance [57], and decomposition [22] could be incorporated into the selection process. Take decomposition, for example. Whenever the re-seed is triggered, solutions in *Archive* should be converted into scalars with a different emphasis from time to time. Then the solution that is closest to the new solution in terms of the scalar value will be selected to compete for acceptance.

## 5.4 Annealing Schedule

The initial temperature is determined in such a way that virtually all transitions are accepted at the beginning 'burn in' period [58]. Let the error be defined as

$$e_r = |(f^* - f_{opt})/f_{opt}| \tag{22}$$

where $f^*$ is the current objective value, and $f_{opt}$ is the desired objective value. The stopping criteria, i.e., the final temperature should be chosen either to control the error defined above. In this research, for cases presented in Sections 6.1 and 6.2 the starting temperature $T_{max}$ and final temperature $T_{min}$ values are set to be 100 and $10^{-4}$. For the test problem analyzed in Section 6.3, the starting temperature $T_{max}$ and final temperature $T_{min}$ values are set to be 1000 and $10^{-2}$, respectively. Another parameter of an annealing schedule is the number of iterations, denoted as *iter*, performed at each temperature. It should be chosen such that the system is sufficiently close to the stationary distribution at that temperature. In this study, *iter* between two consecutive temperature levels are specified as 81, 162 and 200 for cases analyzed in Sections 6.1, 6.2 and 6.3, respectively.

For temperature decrement $T = \Phi(T)$, a common parameter to control the decrement is [59]

$$T_{i+1} = \alpha T_i \tag{23}$$

where $0 < \alpha < 1$ is a constant. Some other cooling schedules available in the literature are logarithmic, Cauchy, and exponential. It has been proved that if annealed sufficiently slowly, SA converges to the global optimum [38]. To have a fair comparison and enable a good exploitation of the solution space, the cooling rate of lowering temperature value is set to be 0.8 for cases analyzed in Sections 6.1 and 6.2, and 0.95 for the test problem analyzed in Section 6.3.

### 5.5 Implementation of Other Algorithms for Comparison

In order to perform fair comparisons, each parameter in AMOSA is set to be the same as that of MOSA/R. We do not execute clustering in AMOSA so that the *Archive* size is unlimited because it is not detrimental to the performance of the algorithm [60]. Furthermore, as mentioned, most solutions along the way are infeasible so that our top priority remains as to find the feasible region, which may be threatened by clustering.

For NSGA-II and MOEA/D, the total number of function evaluations is set in accordance with AMOSA and MOSA/R. Other parameters used follow the original publications of the algorithms [22, 53]. The population size is set to be 300. The distribution indexes in SBX and the polynomial mutation are set to be 20. The crossover rate is 1.00 and the mutation ration is 1/n where n is length of decision vector. In MOEA/D, Tchebycheff approach is used and the size of neighbor population is set to be 20. All initial solutions are generated randomly form the decision space of the problems.

### 6. Case Investigations

In this section, we first examine the performance of MOSA/R using benchmark multi-objective constrained problems as compared to MOEA/D, NAGA-II and AMOSA, and then apply it to the constraint configuration optimization problem defined in Section 2. The simulation results are based on 10 independent test runs. All simulations are carried out within MATLAB on a 2.40GHz Xeon E5620 computer.

The comparison of two single objective optimization (SOO) algorithms is relatively simple because an SOO algorithm only produces one solution. However for multi-objective optimization (MOO), an algorithm will provide a set of solutions that realize the optimal trade-offs between the considered optimization objectives, i.e., Pareto set. Therefore, the comparison of the performance of MOO algorithms has to be based on their Pareto sets. Yet there is no universally accepted definition of optimality in such case. There are multiple optimization goals in our multi-objective optimization: 1) number of feasible solutions; 2) convergence to the true Pareto-optimal set; and 3) maintenance of diversity in solutions of the Pareto-optimal set. Hence, six metrics are briefly explained below, and will be used in Sections 6.1, 6.2 and 6.3 to measure the results in terms of the above-mentioned goals.

*Cardinality (N)*

$N$ is the number of feasible solutions in the Pareto set. For constrained multi-objective optimization problems, finding feasible region is a difficult task of itself. *Cardinality* ($N$) reflects how good the algorithm is in satisfying constraints and exploring feasible search space.

*Inverted Generational Distance (IGD)*

The *IGD* indicator quantifies the degree of convergence by computing the average of the minimum distance of points in the true Pareto front (*PF\**) to points in Pareto front obtained (*PF*), as expressed below,

$$IGD(PF,PF^*) = \frac{\sum_{\mathbf{f}^*\in PF^*,\, i=1}^{|PF^*|} \sqrt{\min_{\mathbf{f}\in PF}\left(\sum_{m=1}^{M}(f_m^i{}^* - f_m)^2\right)}}{|PF^*|} \tag{24}$$

where $M$ is the number of objectives, $f_m$ is the $m$th objective value of $\mathbf{f} \in PF$. In Equation (24), $\min_{\mathbf{f}\in PF}\left(\sum_{m=1}^{M}(f_m^i{}^* - f_m)^2\right)$ calculates the minimum Euclidean distance between the $i$th point in *PF\** and points in *PF*. A lower value of *IGD* indicates better convergence and completeness of the *PF* obtained.

*Hypervolume (HV)*

*HV indicator*

The HV indicator measures convergence as well as diversity. Specifically,

$$HV(PF, r^*) = volume(\bigcup_{x\in PF} v(x, r^*)) \tag{25}$$

where $r^*$ is the reference point which is set to be 1.1 times the upper bound of the Pareto front in the *HV* calculation. The calculation of *HV* requires normalized objective function values. In this research *HV* stands for the percentage covered by the Pareto front of the cuboid defined by the reference point and the origin *(0, 0, 0)*.

*Convergence of two sets (C)*

When the true Pareto-optimal set is not available, we need to determine the relative goodness of two sets of solutions. Let **A**, **B** be two approximation sets consisting of decision vectors. The function *C* maps the ordered pair (**A**, **B**) to the interval [0, 1] in the following manner [23]

$$C(\mathbf{A},\mathbf{B}) = \frac{|\{\mathbf{b}\in \mathbf{B} \mid \exists \mathbf{a}\in \mathbf{A}: \mathbf{a} \prec \mathbf{b}\}|}{|\mathbf{B}|} \tag{26}$$

The value $C(\mathbf{A},\mathbf{B})=1$ means that all solutions in $\mathbf{B}$ are dominated by $\mathbf{A}$. If $C(\mathbf{A},\mathbf{B})=0$ then no point in $\mathbf{B}$ is dominated by points in $\mathbf{A}$. Both directions have to be considered for the comparison since $C(\mathbf{A},\mathbf{B})$ is not necessarily equal to $1-C(\mathbf{B},\mathbf{A})$. As the true Pareto set is unknown, the measure $C$ is used for the performance comparison of two multi-objective algorithms.

We also define that when $\mathbf{B}$ is empty while $\mathbf{A}$ is not, $C(\mathbf{A},\mathbf{B})=1$ and $C(\mathbf{B},\mathbf{A})=0$. When $\mathbf{A}$, $\mathbf{B}$ are both empty, $C(\mathbf{A},\mathbf{B})=C(\mathbf{B},\mathbf{A})=0.5$.

*Minimal Spacing ($S_m$)*

We want to measure the distribution of decision vectors throughout the non-dominated solutions found. Schott [61] proposed a metric measuring the range (distance) variance of neighboring vectors in the Pareto set called *Spacing* (*S*). For a general *M*-dimensional problem, the metric is defined as:

$$S = \sqrt{\frac{1}{N} \sum_{i=1}^{N} (\bar{d} - d_i)^2} \quad (i=1,2,...,N) \tag{27}$$

where

$$d_i = \min_j (\sum_{m=1}^{M} | f_m^i(x) - f_m^j(x) |), i \neq j \tag{28}$$

Then the average $\bar{d}$ of these distances is calculated, and $N$ is the number of solutions on the non-dominated front. A value of zero for this metric indicates all members of the Pareto front currently available are equidistantly spaced. This metric, however, may fail under common scenarios. Thus, we adopt an alternative measure which can overcome the limit of *Spacing* (*S*) called *Minimal Spacing* ($S_m$) as defined in [62].

Initially all solutions are unmarked. We randomly select a solution as seed and mark the seed. We then start computing the nearest distance between the last marked solution and the unmarked solutions. We mark the solution that corresponds to the nearest distance until all solutions are considered and keep track of the sum of the distances. The distance values that correspond to the minimum sum of distances are used as $d_i$ for the computation of the $S_m$, where again Equation (27) is used with $N$ being replaced by $N$-1 since here we have $N$-1 distances. A lower value of $S_m$ indicates better performance of the corresponding MOO technique. When there is only equal or less than one solution found, $S_m$ is set as 1. The range of values of the different objectives are standardized while computing the distances. In other words, when computing $d_i$, the term $|f_m^i(x) - f_m^j(x)|$ is divided by $|F_m^{\max} - F_m^{\min}|$ in order to standardize it, where $F_m^{\max}$ and $F_m^{\min}$ are the maximum and minimum values respectively of the *m*th objective.

*Accounted Proportion P*

In this research, another metric called *Accounted Proportion* is formulated to compare the solutions obtained by several optimization algorithms over multiple simulations at once [66]. Suppose $K$ ($\geq 2$) algorithms are applied to the optimization problem, and the simulation is carried out $L$ times for each algorithm respectively. $PS_i^j$ represents the Pareto set obtained by algorithm $i$ in its $j$th simulation. Let the non-dominated solutions of the union of Pareto sets obtained by algorithm $i$ over $L$ simulations be $PS_i^*$,

$$PS_i^* = \text{Non-dominated}\left(\bigcup_{j \in L}\{PS_i^j\}\right) \tag{29}$$

And the set of non-dominated solutions of all Pareto sets is $PS^*$.

$$PS^* = \text{Non-dominated}\left(\bigcup_{i \in K, j \in L}\{PS_i^j\}\right) \tag{30}$$

Thus for algorithm $i$, the *Accounted Proportion* is

$$P_i = |PS_i^*| / |PS^*| \tag{31}$$

where $|\cdot|$ operation calculates the number of solutions in a set.

### 6.1 Case 1: SRN problem

We first examine the general performance of MOSA/R by applying five algorithms (MOEA/D, NSGA-II, AMOSA, MOSA/R-1.0, MOSA/R-2.0) to a constrained multi-objective benchmark problem SRN [63]:

$$\begin{aligned}
\text{Minimize:} \quad & f_1(\mathbf{x}) = 2 + (x_1 - 2)^2 + (x_2 - 2)^2 \\
& f_2(\mathbf{x}) = 9x_1 - (x_2 - 1)^2 \\
\text{subject to:} \quad & x_1^2 + x_2^2 \leq 225 \\
& x_1 - 3x_2 + 10 \leq 0 \\
& -20 < x_1, x_2 < 20
\end{aligned} \tag{32a-f}$$

The simulation results are presented in Table 5 where three metrics, *Cardinality* and *IGD* and *HV*, are used. The shaded grids in Table 5 indicate the best result in each test in terms of the measures. As demonstrated in Figure 8, MOSA/R-2.0 prevails in *Cardinality* and *IGD*, and is the second best in terms of *HV*. Meanwhile, MOSA/R-1.0 shows better performance than the other three algorithms in *Caridnality* and *IGD*, and reach the third place in *HV* next to NSGA-II and MOSA/R-2.0. In this case, it is relatively easier for each algorithm to find feasible solutions. As shown in Figure 9, all algorithms applied are able to locate the true Pareto front in limited 5,000 function evaluations, but MOSA/R-2.0 exhibits the best coverage over the true Pareto front.

### 6.2 Case 2: TNK problem

Next we apply the algorithms to a more difficult benchmark constrained multi-objective problem TNK [64]. The problem is given as follows:

$$\begin{aligned}
\text{Minimize:} \quad & f_1(\mathbf{x}) = x_1 \\
& f_2(\mathbf{x}) = x_2 \\
\text{Subject to:} \quad & x_1^2 + x_2^2 \geq 1 + 0.1\cos(16\arctan\frac{x_2}{x_1}) \\
& (x_1 - 0.5)^2 + (x_2 - 0.5)^2 \leq 0.5 \\
& 0 < x_1, x_2 < \pi
\end{aligned} \quad (33\text{a-f})$$

We modify Equation (33f) by enlarging the design space to $0 < x_1, x_2 < 100$ in order to make it more challenging in finding feasible solutions. Compared to SRN, it is harder to locate the feasible region because this problem has larger decision space and the true Pareto front is discrete and non-linear. For each algorithm, it runs until 10,000 function evaluations are reached. The simulation results are presented in Table 5. As demonstrated in Figure 8, three Simulated Annealing based algorithms have better performance than Genetic algorithms, which validates our discussion in Section 4. MOSA/R-2.0 has the best overall performance among the five algorithms. Consistent with the case evaluated in Section 6.1, both MOSA/R algorithms achieve more feasible solutions than MOEA/D, NSGA-II and AMOSA. In addition, as Figure 9 illustrates, both MOSA/R maintain relatively good diversity compared to other algorithms.

In both cases illustrated in Sections 6.1 and 6.2, MOSA/R has an edge over some other contemporary approaches when applied to constrained multi-objective benchmark problems. Given limited number of function evaluations, MOSA/R converges faster and yields more useful feasible results.

### 6.3 Case 3: Configuration optimization test problem

In this sub-section, we will solve the configuration optimization test problem defined in Section 2. For this problem, there are three main goals that an MOO algorithm should reach. It should find as many feasible solutions as possible. It should converge as close to the true Pareto front as possible. It should also maintain as diverse a solution set as possible. As mentioned earlier in Section 6, the comparison metrics $N$ shows the number of feasible solutions (the larger the better), $C$ represents the degree of convergence with respect to the compared set of solution (the closer to 1 the better), $P$ illustrates how well the algorithm converges from a different perspective when the strategies applied are considered all at once (the larger the better), and $S_m$ indicates the diversity (the smaller the better).

The numerical experiment is designed in such a way that we decrease one parameter *SL* (i.e., tightening the constraints), which is the length of cube side, from 9.4 to 8.2 with a step of size 0.1. The cut-off *SL* value is chosen to be 8.2 because this is when AMOSA have difficulties finding feasible solutions. For each *SL* value, we apply AMOSA, MOSA/R-1.0 and MOSA/R-2.0. All algorithms are executed 10 times, and the results reported are the mean values and standard deviations obtained over 10 runs. Genetic algorithms (MOEA/D and NSGA-II) are not included in the comparison because they are not able to find feasible solutions. The possible reasons have been discussed in Section 4. Notations AM and MR in the tables mentioned below stand for AMOSA and MOSA/R respectively.

Tables 6 and 7 demonstrate the relative performance of AMOSA, MOSA/R-1.0, and MOSA/R-2.0 in terms of the three metrics *N*, *C* and $S_m$. The comparisons are also illustrated in Figures 10 and 11. The number of feasible solutions (*Cardinality*) found by MOSA/R-1.0 and MOSA/R-2.0 are always greater than that of AMOSA, as can be seen from Figure 10(a), and MOSA/R-2.0 has a clear advantage in this contest because the re-seed scheme of MOSA/R-2.0 is designed for finding feasible solutions under hard constraints. When AMOSA and MOSA/R-1.0 start to show no feasible solution as constraints are tightening, MOSA/R-2.0 excels at *Cardinality* considerably.

MOSA/R-2.0 is the best among three in the regard of *Convergence* as well. Meanwhile, MOSA/R-1.0 provides generally better performance than AMOSA. The exception happens when *SL* equals to 8.4 (Figure 10(d)) which may not be the case when more simulations are executed. The reason for such fluctuation is that all three algorithms are heuristic that do not guarantee to find the exact optimal solutions. However, they are capable of providing immediate results that are very close to optimality efficiently and handle optimization problems that are currently out of the reach of theoretically rigorous methodology [65]. As a result, the quality of solutions of heuristic methods has no robust justification. The *Minimal Spacing* curve of MOSA/R-2.0 remains low, indicating that it maintains better diversity than MOSA/R-1.0 and AMOSA. MOSA/R-1.0 is no worse than AMOSA at *Minimal Spacing* in all tested *SL*s. We should note here that for the test problem, a set of well-spread solutions doesn't necessary mean it is superior to that with relatively bad diversity. Nevertheless, because a set of well-spread solutions usually belong to the same class of configuration, a set of ill-distributed solutions may belong to different classes of configuration meaning possibly better diversity.

Figure 12 depicts the best solutions found in 10 test runs by the three algorithms respectively as *SL* varies from 9.4 to 8.2. Figure 13 shows the combined best Pareto set over 10 runs for each *SL*. The *Accounted Proportion* for each algorithm is calculated and shown in Table 8. Once again, MOSA/R-2.0 prevails, and MOSA/R-1.0 is the second best. Figure 14 depicts average *Accounted Proportion* of each algorithm considering all simulations.

The numerical results illustrate that MOSA/R has desirable overall performance in terms of *Cardinality*, *Convergence*, *Minimal Spacing,* and *Accounted Proportion* for the test problem as the constrained level elevates. To be more specific, MOSA/R-2.0 is far ahead of AMOSA in all four contests while MOSA/R-1.0, which could be considered as a general version of MOSA/R-2.0 or an improved version of AMOSA, has an edge over AMOSA. In this research, due to length limitations, only the interval where AMOSA starts having difficulties finding feasible solutions is considered in order to demonstrate the capability of MOSA/R under such situation. Videos that illustrate the optimization process of MOSA/R-1.0 and MOSA/R-2.0 applied to the configuration optimization test problem when *SL* is 8.7 can be found in https://youtu.be/zmdwNsyZYow and https://youtu.be/YeHrtSHY5ss respectively. As demonstrated in the first video, MOSA/R-1.0 locates the first feasible solution at temperature 32.1723 and that is the only feasible solution it is capable of finding. On the other hand, as shown in the second video, MOSA/R-2.0 finds the first feasible solution at temperature 440.1267, and finally provides the user with 32 feasible solutions. Notice in the second video, the current solution frequently jumps back to the archived solutions that when re-seed happens.

## 7. Conclusion

This research provides a systematic framework of solving configuration optimization problems with multiple hard constraints. For this class of problems, the optimizer is required not only to locate the feasible regions, but also to maintain convergence and diversity among the feasible solutions. We develop an improved MOSA algorithm (MOSA/R) towards MOO problems, featuring a newly designed re-seed scheme that is able to redirect the optimizer towards the regions of better feasibility and optimality. Two versions of MOSA/R are formulated, i.e., the general version MOSA/R-1.0 and a version for constrained problems MOSA/R-2.0. For benchmark constrained MOO problems, MOSA/R shows better performance than MOEA/D, NSGA-II and AMOSA. For the constrained configuration optimization problem, it is identified that when all methods are applicable, MOSA/R-2.0 yields better results than MOSA-1.0 and AMOSA, and MOSA/R-1.0 has better performance than AMOSA in general senses. This new framework can be employed to facilitate an integrated process of design automation and manufacturing optimization, thereby benefiting both existing design/manufacturing practices and future additive manufacturing.

Table 1 Specifications of cylinders

| Component | Diameter $2r$ (inch) | Length $l$ (inch) |
|---|---|---|
| Cylinder 1 | 1.25 | 5 |
| Cylinder 2 | 1.25 | 5 |
| Cylinder 3 | 1.00 | 4 |
| Cylinder 4 | 1.00 | 4 |
| Cylinder 5 | 1.00 | 4 |
| Cylinder 6 | 0.75 | 3 |

Table 2 Connectivity relation between cylinders

| Connective Line | Start Point | End Point |
|---|---|---|
| 1 | Window 2 of Cylinder 1 | Window 1 of Cylinder 6 |
| 2 | Window 2 of Cylinder 6 | Window 1 of Cylinder 2 |
| 3 | Window 2 of Cylinder 2 | Window 1 of Cylinder 4 |
| 4 | Window 2 of Cylinder 4 | Window 1 of Cylinder 3 |
| 5 | Window 2 of Cylinder 3 | Window 1 of Cylinder 5 |
| 6 | Window 2 of Cylinder 5 | Window 1 of Cylinder 1 |

Table 3 Domination relations

| Relation | Symbol | Interpretation in objective space |
|---|---|---|
| **a** dominates **b** | $a \prec b$ | **a** is not worse than **b** in all objectives and better in at least one objective |
| **b** dominates **a** | $b \prec a$ | **b** is not worse than **a** in all objectives and better in at least one objective |
| Non-dominant to each other | $b \cong a$ | **a** is worse than **b** in some objectives but better in some other objectives |

Table 4 Decision procedures of MOSA/R as compared to AMOSA

| | | | MOSA/R | AMOSA |
|---|---|---|---|---|
| $New \prec Archive$ /*Case 1*/ | $New \prec Cur$ | | *Accept* and *Update* | |
| | $New \cong Cur$ | | *Accept* and *Update* | |
| | $New \succ Cur$ | | Not possible | |
| $New \succ Archive$ /*Case 2*/ | $New \succ Cur$ /*2a*/ | $Cur \in Archive$ /*2a-1*/ | *Accept* with *prob* | *Accept* with *prob* |
| | | $Cur \notin Archive$ /*2a-2*/ | **Re-seed** with *prob* *Accept* with (1- *prob*)\**prob'* | |
| | $New \prec Cur$ /*2b*/ | | *Accept* | **Re-seed** with *prob* *Accept* with (1- *prob*) |
| | $New \cong Cur$ /*2b*/ | | *Accept* with *prob* | |
| $New \cong Archive$ /*Case 3*/ | $New \prec Cur$ | | *Accept* and *Update* | |
| | $New \cong Cur$ | | *Accept* and *Update* | |
| | $New \succ Cur$ | | Not possible | |

Table 5 *Cardinality*, *IGD* and *HV* measures on the constrained benchmark tests SRN and TNK over 10 runs in terms of mean

|  | SRN | | | TNK | | |
| --- | --- | --- | --- | --- | --- | --- |
|  | *Cardinality* | *IGD* | *HV* | *Cardinality* | *IGD* | *HV* |
| MOEA/D | 59.6667 | 2.9692 | 0.9318 | 5.3333 | 0.14838 | 0.25995 |
| NSGA-II | 132.5 | 1.6596 | 0.9519 | 32.5 | 0.1179 | 0.3097 |
| AMOSA | 24 | 9.89 | 0.93155 | 47.16667 | 0.010917 | 0.38555 |
| MOSA/R-1.0 | 214.3333 | 3.2863 | 0.9325 | 49.5 | 0.0113 | 0.3861 |
| MOSA/R-2.0 | 450.1667 | 0.7966 | 0.9476 | 53.1667 | 0.010633 | 0.3869 |

Table 6 *Cardinality* and *Minimal Spacing* measures on the configuration optimization test problem over 10 runs in terms of mean and standard deviation

| SL | *Cardinality (N)* | | | *Minimal Spacing ($S_m$)* | | |
| --- | --- | --- | --- | --- | --- | --- |
|  | AM | MR-1.0 | MR-2.0 | AM | MR-1.0 | MR-2.0 |
| 9.4 | 6.3 (2.87) | 8.7 (3.37) | 12.3 (5.23) | 0.3756 (0.23) | 0.2822 (0.13) | 0.1951 (0.08) |
| 9.3 | 5.7 (2.40) | 6.1 (2.92) | 16.5 (6.93) | 0.3368 (0.28) | 0.3177 (0.26) | 0.1003 (0.04) |
| 9.2 | 5.5 (3.54) | 9.7 (4.06) | 17.7 (7.73) | 0.4279 (0.40) | 0.2298 (0.11) | 0.1218 (0.06) |
| 9.1 | 8.0 (4.71) | 7.1 (3.73) | 16.0 (8.65) | 0.3471 (0.35) | 0.2693 (0.14) | 0.1398 (0.13) |
| 9.0 | 4.0 (2.40) | 5.0 (3.09) | 16.8 (6.66) | 0.4946 (0.36) | 0.4359 (0.25) | 0.1178 (0.05) |
| 8.9 | 3.5 (1.35) | 6.3 (2.63) | 16.6 (5.60) | 0.5618 (0.33) | 0.3182 (0.26) | 0.1078 (0.04) |
| 8.8 | 2.9 (2.56) | 3.1 (1.79) | 16.6 (6.02) | 0.6469 (0.33) | 0.6457 (0.39) | 0.0994 (0.04) |
| 8.7 | 2.5 (2.68) | 5.0 (3.53) | 19.2 (8.35) | 0.7693 (0.37) | 0.5301 (0.34) | 0.1041 (0.05) |
| 8.6 | 3.3 (2.26) | 5.0 (3.56) | 11.7 (8.67) | 0.6990 (0.27) | 0.5112 (0.35) | 0.2526 (0.28) |
| 8.5 | 1.9 (1.20) | 3.6 (2.99) | 16.2 (8.04) | 0.8069 (0.31) | 0.6194 (0.36) | 0.1099 (0.06) |
| 8.4 | 1.9 (2.18) | 3.8 (5.01) | 19.4 (10.85) | 0.7602 (0.39) | 0.6384 (0.40) | 0.0900 (0.05) |
| 8.3 | 1.1 (1.85) | 1.4 (1.35) | 17.9 (9.16) | 0.9199 (0.25) | 0.8237 (0.29) | 0.1181 (0.13) |
| 8.2 | 0.6 (1.08) | 1.2 (2.15) | 19.6 (7.30) | 0.9023 (0.31) | 0.8458 (0.33) | 0.0798 (0.03) |

Table 7 *Convergence* measure on the configuration optimization test problem over 10 runs in terms of mean and standard deviation

| SL | *Convergence (C)* | | | | | |
| --- | --- | --- | --- | --- | --- | --- |
|  | C(MR1,AM) | C(AM,MR1) | C(MR1,MR2) | C(MR2,MR1) | C(MR2,AM) | C(AM,MR2) |
| 9.4 | 0.5985 (0.43) | 0.2279 (0.34) | 0.2474 (0.42) | 0.4303 (0.45) | 0.6845 (0.39) | 0.1000 (0.32) |
| 9.3 | 0.5525 (0.35) | 0.1625 (0.35) | 0.0633 (0.13) | 0.6848 (0.40) | 0.8092 (0.28) | 0.0143 (0.05) |
| 9.2 | 0.5443 (0.49) | 0.2883 (0.41) | 0.2042 (0.36) | 0.5533 (0.48) | 0.8667 (0.32) | 0.0765 (0.24) |
| 9.1 | 0.3980 (0.37) | 0.2387 (0.35) | 0 (0) | 0.7031 (0.30) | 0.7317 (0.38) | 0.1000 (0.32) |
| 9.0 | 0.4500 (0.48) | 0.2857 (0.46) | 0.1209 (0.16) | 0.5307 (0.43) | 0.5958 (0.47) | 0.0944 (0.20) |
| 8.9 | 0.4850 (0.45) | 0.0433 (0.11) | 0.0545 (0.12) | 0.8494 (0.29) | 0.9000 (0.32) | 0.0241 (0.08) |
| 8.8 | 0.4333 (0.50) | 0.3400 (0.47) | 0.0600 (0.19) | 0.6150 (0.46) | 0.7333 (0.41) | 0 (0) |
| 8.7 | 0.5972 (0.45) | 0.2500 (0.42) | 0.1597 (0.32) | 0.7067 (0.42) | 0.8000 (0.42) | 0.0580 (0.14) |
| 8.6 | 0.4226 (0.46) | 0.3200 (0.42) | 0 (0) | 0.5000 (0.48) | 0.9333 (0.14) | 0 (0) |
| 8.5 | 0.6500 (0.47) | 0.3000 (0.48) | 0.1352 (0.24) | 0.4250 (0.50) | 0.7667 (0.42) | 0 (0) |
| 8.4 | 0.3700 (0.40) | 0.3733 (0.40) | 0 (0) | 0.7833 (0.34) | 0.8000 (0.42) | 0 (0) |
| 8.3 | 0.6000 (0.46) | 0.3500 (0.41) | 0 (0) | 0.8500 (0.34) | 0.9833 (0.05) | 0 (0) |
| 8.2 | 0.5500 (0.37) | 0.4500 (0.37) | 0 (0) | 0.9583 (0.09) | 1.0000 (0) | 0 (0) |

Table 8 *Accounted Proportion* of solutions obtained by each algorithm in the combined Pareto sets

| SL | *Accounted Proportion (P)* | | |
|---|---|---|---|
| | AMOSA | MOSA/R-1.0 | MOSA/R-2.0 |
| *9.4* | 0% | 16.00% | 84.00% |
| *9.3* | 0% | 2.86% | 97.14% |
| *9.2* | 0% | 0% | 100% |
| *9.1* | 2.63% | 5.26% | 92.11% |
| *9.0* | 0% | 22.22% | 77.78% |
| *8.9* | 0% | 0% | 100% |
| *8.8* | 0% | 0% | 100% |
| *8.7* | 0% | 21.88% | 78.13% |
| *8.6* | 0% | 38.10% | 61.90% |
| *8.5* | 0% | 4.17% | 95.83% |
| *8.4* | 0% | 0% | 100% |
| *8.3* | 0% | 7.14% | 92.86% |
| *8.2* | 0% | 0% | 100% |
| *Average* | 0.2023% | 9.0485% | 90.7500% |

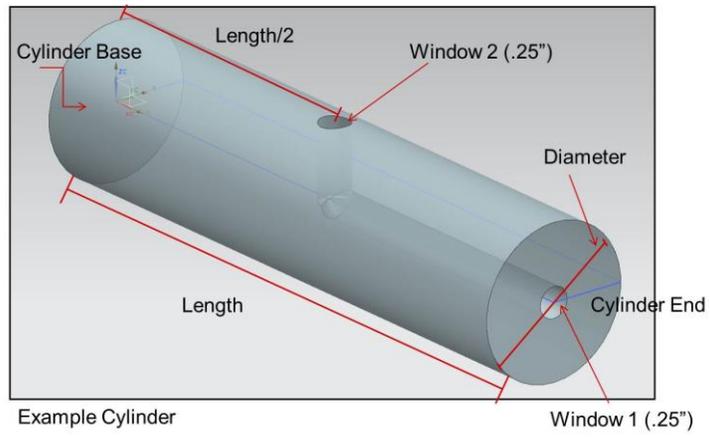
Figure 1 Cylinder example.

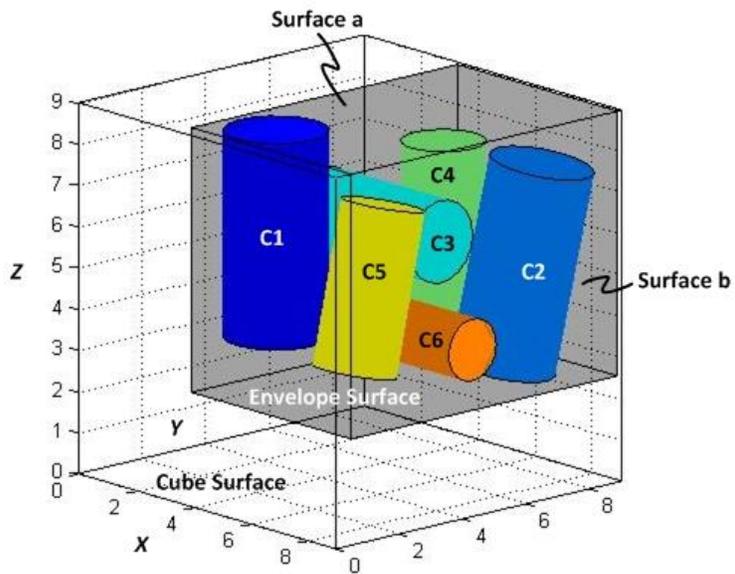
Figure 2 Example configuration

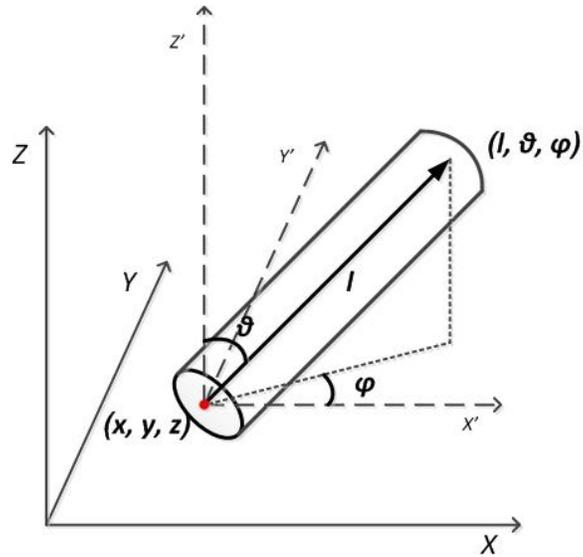

Figure 3 Cylinder in coordinate system.

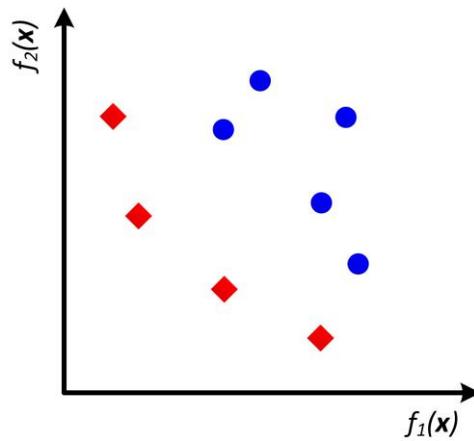

Figure 4 Domination relations within a set of solutions for a two objectives minimization problem (◆: Non-dominated solutions; ●: Dominated solutions).

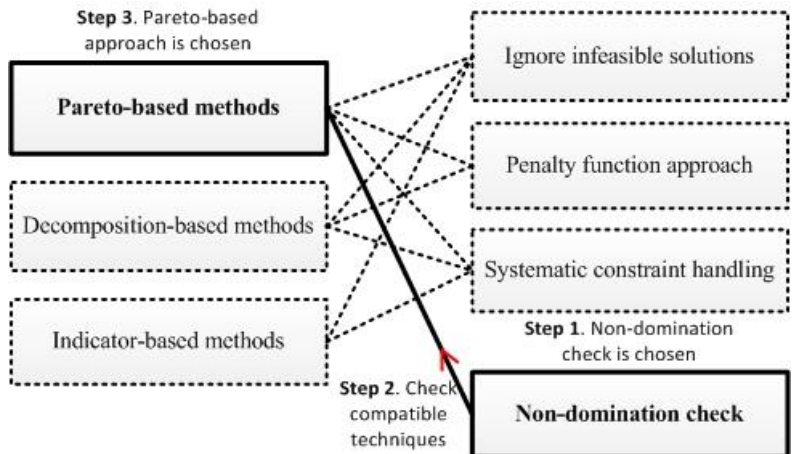

Figure 5 Compatibility relationships between multi-objective handling and constraint handling techniques.

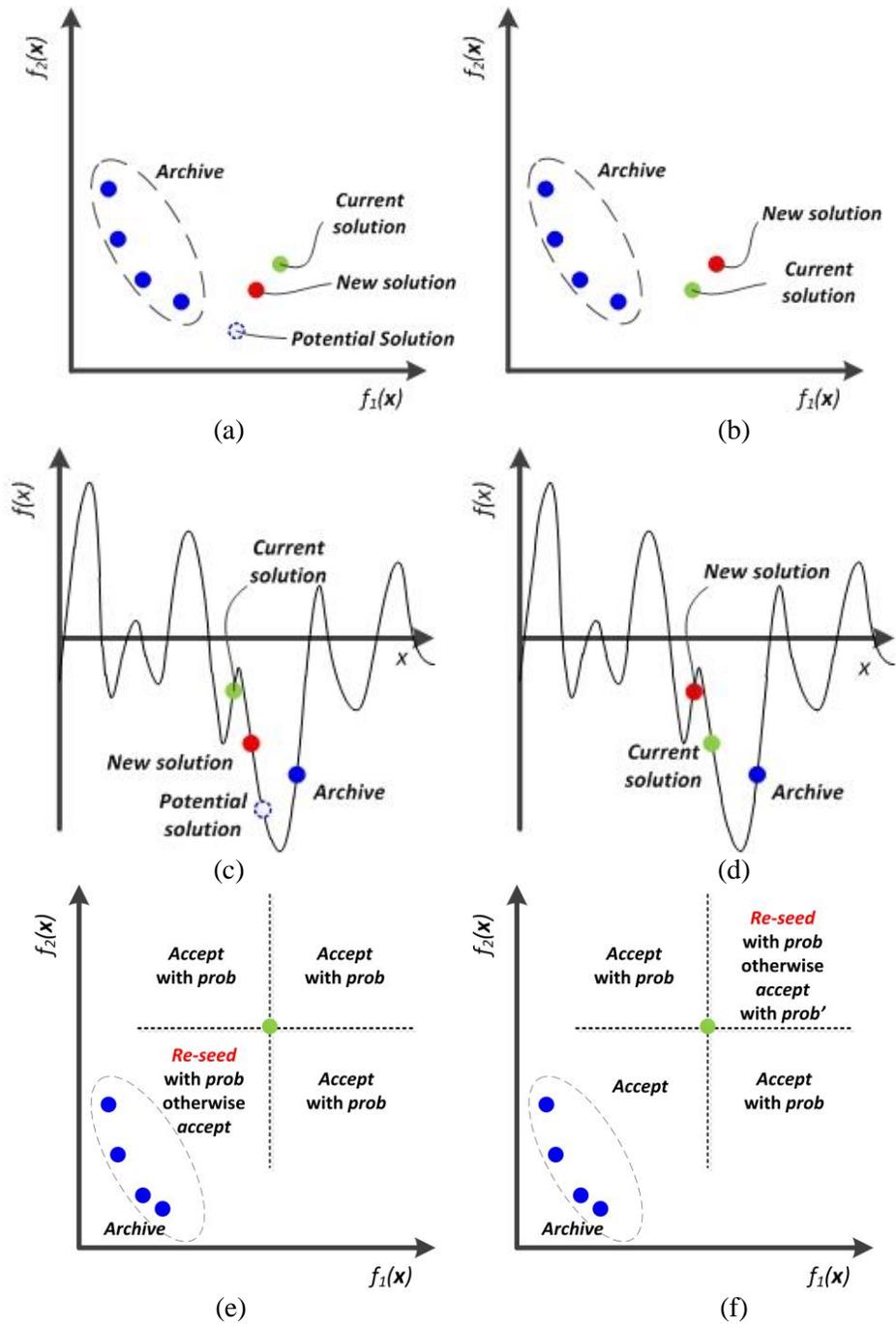

Figure 6 Re-seed timing comparisons of AMOSA (a) (c) (e) and MOSA/R (b) (d) (f).

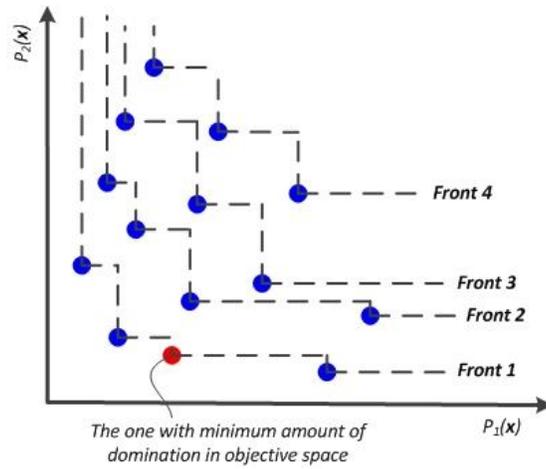

Figure 7 How to re-seed when constraints come into effect.

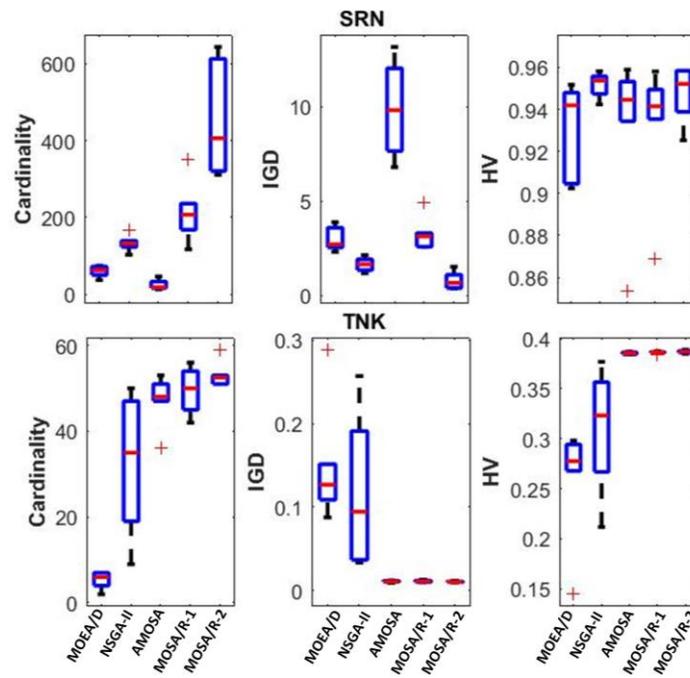

Figure 8 Figure 8 Box plots of five algorithms w.r.t. *Cardinality*, *IGD* and *HV* on SRN and TNK
(From left to right: MOEA/D, NAGA-II, AMOSA, MOSA/R-1.0, and MOSA/R-2.0)

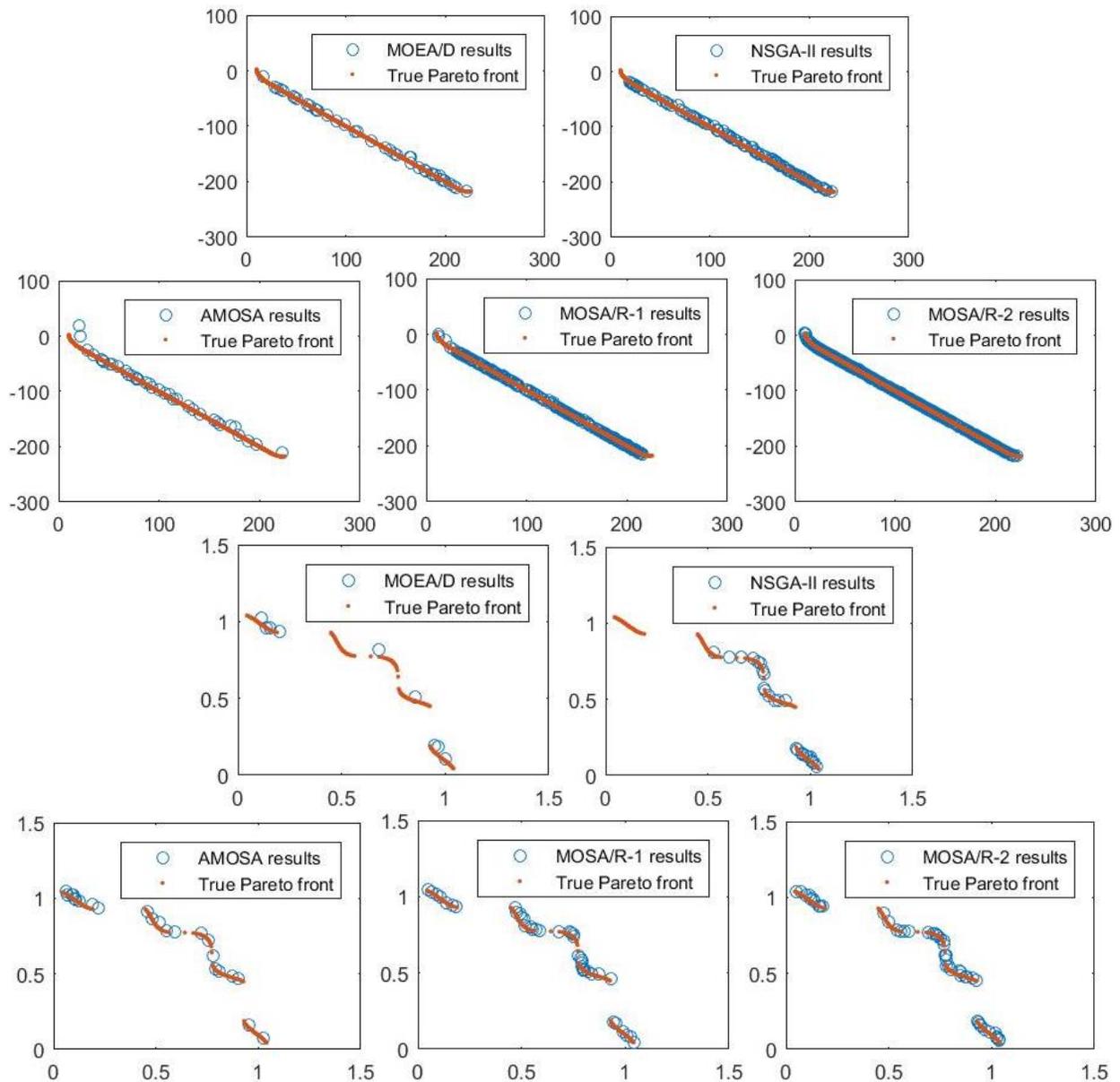
Figure 9 Pareto front obtained by each algorithm for test instances SRN and TNK
(First 5 SRN, last 5 TNK)

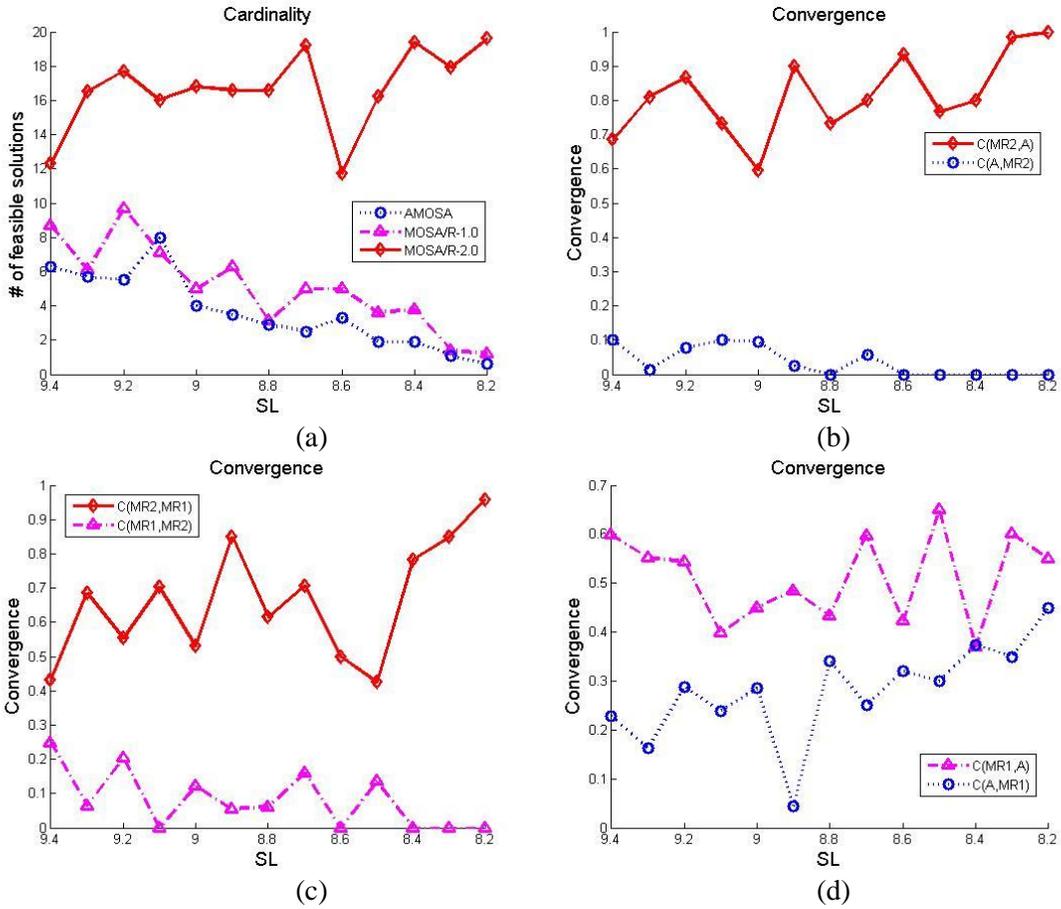

Figure 10 *Cardinality* and *Convergence* with *SL* varying from 9.4 to 8.2.

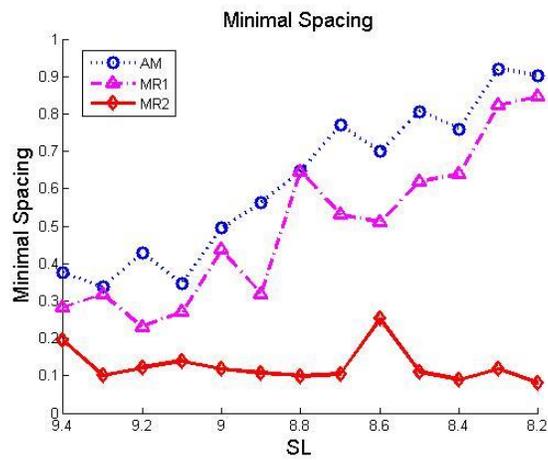

Figure 11 *Minimal Spacing* with *SL* varying from 9.4 to 8.2.

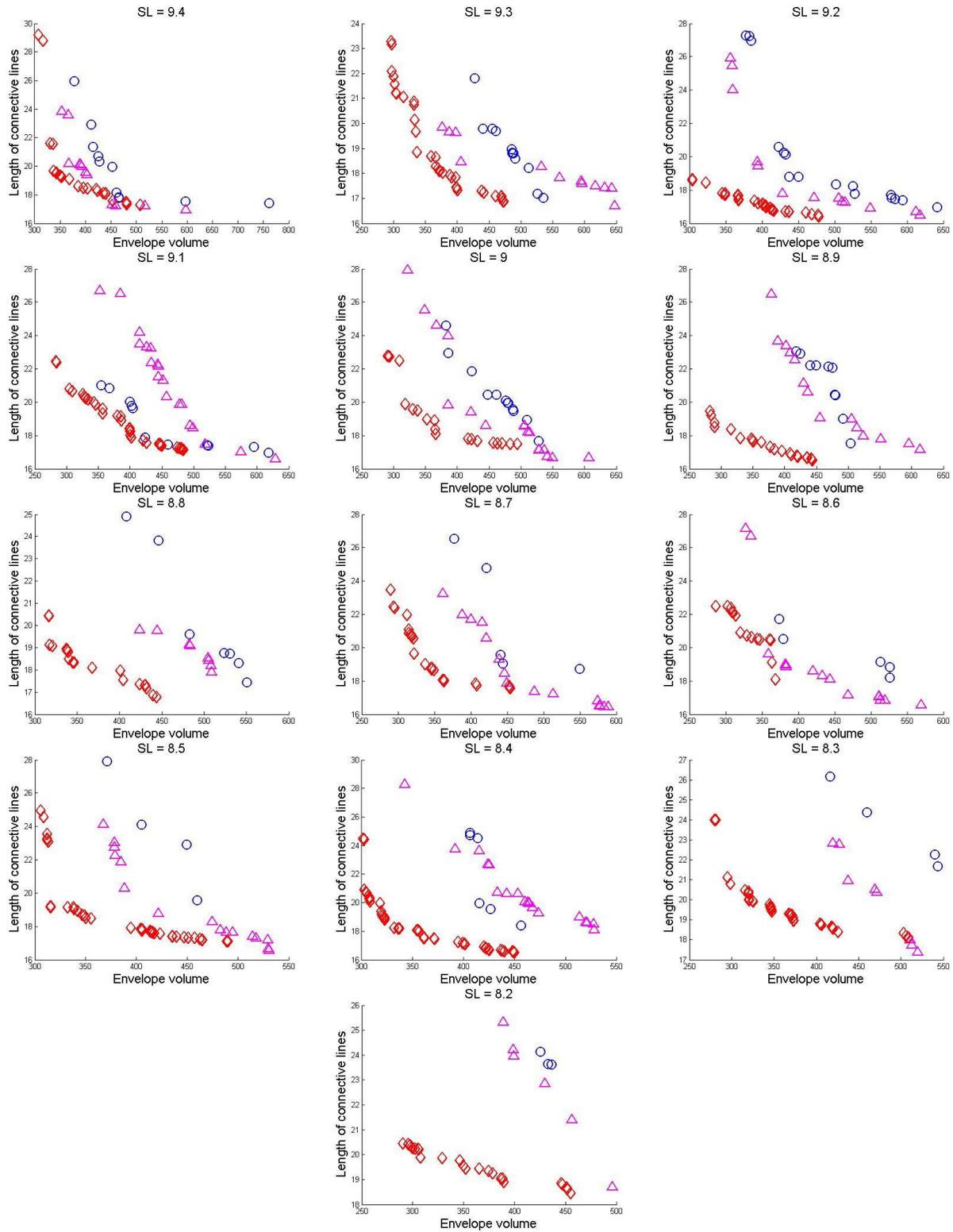

Figure 12 Pareto set after 10 test runs of the three algorithms as *SL* varying from 9.4 to 8.2
(◇: MOSA/R-2.0; △: MOSA/R-1.0; O: AMOSA).

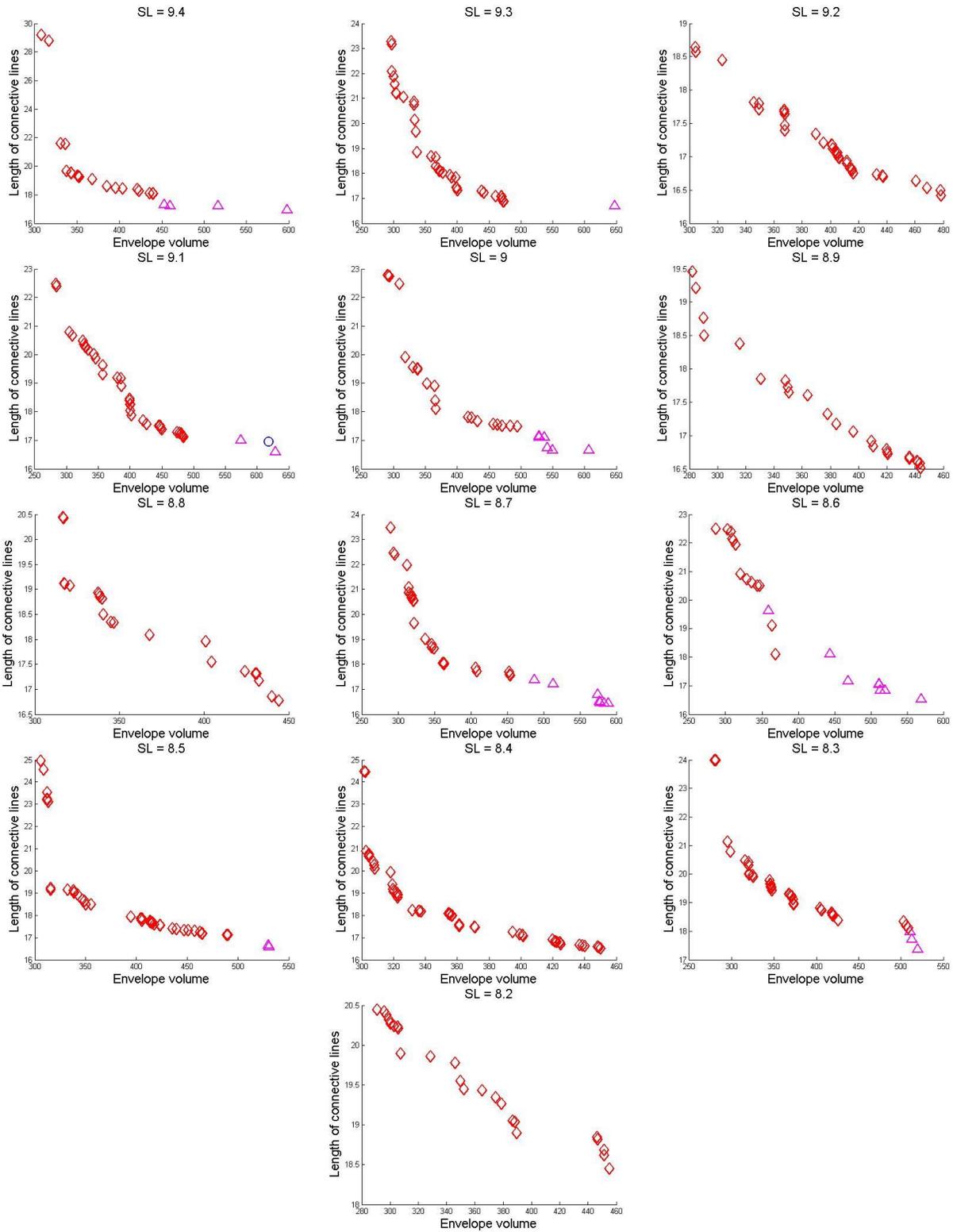

Figure 13 Best combined Pareto set over 10 runs as *SL* varying from 9.4 to 8.2
(◇: MOSA/R-2.0; △: MOSA/R-1.0; ○: AMOSA).

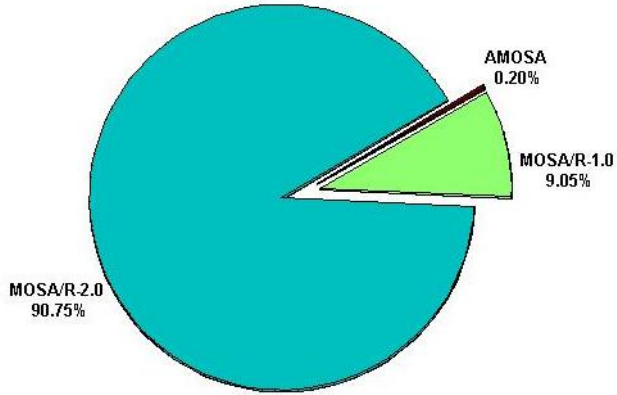

Figure 14 Average *Accounted Proportion* of each algorithm.